\documentclass[twocolumn,showkeys,showpacs,preprintnumbers,prd,superscriptaddress,nofootinbib]{revtex4-1}
\usepackage{graphicx}	
\usepackage{amssymb}
\usepackage{dcolumn}
\usepackage{mathtools}
\usepackage{amsmath}
\usepackage{xcolor}
\usepackage{color}
\usepackage[normalem]{ulem}
\usepackage{theorem}
\usepackage{subfigure, rotating, bm, array}
\usepackage[pagebackref=false, colorlinks=true]{hyperref}
\hypersetup{linkcolor=blue, citecolor=blue,urlcolor=blue} 

\begin{document}

\title{Relativistic time delay analysis of pulsar signals near ultra-compact objects}

\author{Viraj Kalsariya}
\email{virajrk6@gmail.com}
\affiliation{International Center for Space and Cosmology, Ahmedabad University, Ahmedabad 380009, Gujarat, India}
\author{Parth Bambhaniya}%
\email{grcollapse@gmail.com}
\affiliation{International Center for Space and Cosmology, Ahmedabad University, Ahmedabad 380009, Gujarat, India}
 \author{Pankaj S. Joshi}%
 \email{psjcosmos@gmail.com}
 \affiliation{International Center for Space and Cosmology, Ahmedabad University, Ahmedabad 380009, Gujarat, India}

\begin{abstract}
The upcoming discoveries of pulsars orbiting the center of the Milky Way will present unparalleled opportunities to examine the causal structure of the spacetime geometry of Sagittarius A*. In this paper, we investigate the fully relativistic propagation time delay of pulsar signals in the Joshi-Malafarina-Narayan (JMN-1) and Janis-Newman-Winicour (JNW) spacetimes. This delay arises from the spacetime curvatures in the vicinity of these ultra-compact objects, induced by the intense gravitational field near the Galactic Center (GC). Using the principles of gravitational lensing, we compute the arrival time of photons originating from a pulsar in orbit around the GC.
To validate our approach, we compare our time delay analysis of the Schwarzschild black hole with the corresponding delay in the post-Newtonian framework. Subsequently, we find that the propagation time of pulsar signal is greater and lesser for the given horizon-less ultra-compact objects for direct and indirect propagation respectively. Therefore, our results suggest quite significant propagation time delay differences in JMN-1 and JNW spacetimes, when compared to the Schwarzschild black hole case. This can be inferred as a possible distinguishing feature for these ultra-compact objects' geometries.


\end{abstract}

\maketitle


\section{Introduction}

In 1939, Oppenheimer, Snyder, and Datt (OSD) introduced a collapse model, which is a solution for the dynamical collapse of a spherically symmetric, homogeneous dust cloud \cite{Oppenheimer:1939ue,Datt:1938}. This model assumed ideal conditions, such as uniform density within the collapsing star and zero gas pressure, simplifying the Einstein equations. This model hinted at the black hole formation as the final fate of continued collapse, with a spacetime singularity at the center, hidden within an event horizon. Based on the OSD model, later in 1969, Penrose proposed the Cosmic Censorship Conjecture (CCC) suggesting that space-time singularities must remain hidden from distant observers when a massive star gravitationally collapses \cite{penrose}. 

However, due to highly ideal assumptions used in the OSD model, the question of the final fate of collapse remained open under physically more realistic conditions. While the CCC implies that gravitational collapse must lead exclusively to black holes only, subsequent research highlighted that an inhomogeneous matter collapse, for example with density higher at the center and decreasing slowly as we move away, can result in naked singularities \cite{psj1,psj2,jnw,bst}. In this latter case, the spacetime singularity forming as a collapse final state is no longer hidden within an event horizon of gravity. This highlights the importance of considering the role of various initial conditions from which the gravitational collapse evolves, as these dictate the formation of singularities with or without event horizons \cite{psj2}.

Now a crucial question emerges: Would a massive collapsing star conclude as a black hole or a naked singularity at the end of its life cycle? This issue poses an ongoing challenge. To address this, understanding the key physical and geometrical attributes of black holes and their alternatives is essential. As highlighted earlier, relatively realistic astrophysical scenarios can yield naked singularities, which may have substantial implications for our comprehension of the universe.

In principle, the theory of general relativity (GR) predicts that the spacetime singularity forms necessarily when large enough mass collapses under its own gravity. Nevertheless, it does not necessarily simultaneously enforce the formation (or otherwise) of an event horizon.  In \cite{psj2}, Joshi, Malafarina, and Narayan have shown that a non-zero tangential pressure can prevent the formation of trapped surfaces around the core of a high-density region of a collapsing matter cloud, resulting in a central naked singularity in large co-moving time. All the same, Janis, Newman, and Winicour obtained a minimally coupled mass-less scalar field solution of the Einstein field equations \cite{jnw}. The subsequent observational aspects have been studied in JMN-1 and JNW spacetimes including shadows and accretion disk properties \cite{Saurabh:2023otl,Saurabh:2022jjv,shaikh1,gyulchev,Solanki:2021mkt,Sau:2020xau,Chowdhury:2011aa}, relativistic orbits of S2 star \cite{Bambhaniya:2022xbz,Bambhaniya:2019pbr,Joshi:2019rdo,Dey:2019fpv}, hot spot \cite{Chen:2023knf}, energy extraction \cite{Patel:2022jbk,Patel:2023efv,Acharya:2023vlv}, echos and Quasi-Normal modes \cite{Chowdhury:2020rfj,Stashko:2023ffs}, etc.

Recently, the Event Horizon Telescope (EHT) collaboration has announced a major breakthrough in the imaging of the ultra-compact object at the center of our galaxy \cite{EventHorizonTelescope:2022xnr}. 
The EHT collaboration made remarkable progress in exploring the evidence for the event horizon of a black hole, significantly advancing our understanding and opening new avenues for other possibilities. All the same, based on the results and analyses presented in the EHT papers (see e.g.  \cite{EventHorizonTelescope:2022xqj}), they have importantly suggested that the observational evidence does not definitively rule out the possibility that Sgr A* could be a naked singularity. Thus the EHT team has recognized this possibility, favoring an open and careful interpretation of their findings. The accretion flows onto these singularities have spectra and images nearly identical to those of a Schwarzschild black hole, therefore indicating that a JMN-1 naked singularity with a photon sphere may be one of the best possible black hole mimickers for Sgr A* \cite{EventHorizonTelescope:2022xqj}. 
Similarly, the shadows created by compact objects, including black holes, naked singularities, and also gravastars and wormholes, have received an attention in such a context \cite{shaikh1,gyulchev, ohgami_2015,Sakai,Bambhaniya:2021ugr,Vertogradov:2024fva,Bambhaniya:2021ybs,ABJoshi}.

From such a perspective, since other compact objects do cast similar shadows as a black hole (see also \cite{Vagnozzi:2022moj} and references therein), the JMN-1 and JNW spacetimes with a photon sphere appear to be best possible black hole mimickers for Sgr A* as mentioned above \cite{psj2,shaikh1, Saurabh:2022jjv, Saurabh:2023otl}. 
The point is, while there is strong evidence that there is a huge concentration of mass in the center of our Milky Way galaxy, the question of whether or not it is a black hole with an event horizon is still open and remains to be examined carefully. 
These results have piqued an interest in the analysis of the nature of the object Sgr A*, and it is intriguing to ask and to determine whether it is a supermassive black hole (SMBH) or possibly an ultra-compact object without a horizon or a naked singularity.

Within such a context, in this paper, we have therefore investigated the propagation time delay of pulsar signals in the vicinity of naked singularities. The potential identification of pulsars in orbit around the ultra-compact object at the center of the Milky Way provides a unique opportunity to test GR. Along with the timing analysis of the radio pulses emitted by these pulsars offers distinctive avenues for exploring the gravitational dynamics of the compact objects  \cite{Wex:1998wt, Pfahl_2004, Kramer:2004hd, Liu:2011ae, Psaltis:2015uza}. The implications of these observations on our current understanding of gravity are unparalleled, providing a unique opportunity to examine GR in the strong-field regime with exceptional precision. Additionally, it encourages rigorous investigation against alternative compact objects with remarkable accuracy. Because of this interesting potential and possibility to resolve many related mysteries, a large number of radio pulsar searches had begun near the GC of the Milky Way galaxy \cite{10.1093/mnras/274.1.L43, 10.1111/j.1745-3933.2006.00232.x, Deneva_2009, 10.1111/j.1365-2966.2010.17790.x}. Also, recent research suggests that there should be around 100 to 1000 pulsars orbiting within the range of 100 years of orbital period with an average of 100 pulsars in the range of 10 years orbit \cite{Pfahl_2004, Zhang_2014, 10.1093/mnras/stx1661, 10.1093/mnrasl/slu025}. This directs and motivates the search of pulsars in tight orbits around Sgr A*.

Even with considerable efforts, only six pulsars were detected within 15 arc-minutes of Sgr A* \cite{Deneva_2009}, and just one radio magnetar was detected at an angular distance of 2.4 arc-seconds from Sgr A* \cite{Kennea_2013, Mori_2013, Rea_2013}. Due to the interstellar scattering process, which affects the temporal broadening of the pulses in a heavily turbulent and ionized interstellar medium at the GC \cite{2002astro.ph..7156C}, it is extremely hard to detect pulsars in tight orbit at the GC. The effectiveness of this procedure significantly relies on the observing frequency $\nu$ ($\propto \nu^4$), rendering the conventional periodicity search methods at frequencies around $\nu \sim 10 GHz$ largely ineffective, even for pulsars with extended periods. The widening of pulse duration due to temporal broadening cannot be rectified or counteracted through instrumental adjustments or corrections \cite{2013IAUS..291..382E}. To address this challenge, one potential approach is to shift pulsar searches to higher observing frequencies to reduce the scattering effect. However, considering the nature of pulsar spectra, which decrease flux density ($\propto \nu^\alpha$) as frequency increases (for $\alpha<0$ as shown in \cite{Wharton_2012}), higher frequencies correspond to fainter intrinsic source brightness. 

Consequently, despite efforts, recent high-frequency pulsar searches in the GC, even using wavelengths of 2 and 3 mm  \cite{2021A&A...650A..95T}, have failed to detect new pulsars in that area. Currently, new generation telescopes such as the Square Kilometre Arry (SKA) \cite{2015aska.confE..40K}, Five-hundred-meter Aperture Spherical Telescope (FAST) \cite{2011IJMPD..20..989N}, the Event Horizon Telescope (EHT) \cite{EHT:2023hcj} are actively searching for pulsars near Sgr A* and conducting time delay analyses. Hence, studying the time delay of light pulsations emitted by pulsars orbiting near the Milky Way GC is highly beneficial. Subsequently, we can evaluate how these time delays differ between black hole and horizon-less singularity spacetime models, aiding in the identification of the central compact object. This analysis will help us distinguish the Schwarzschild Black hole from JMN-1 and JNW naked singularities.

The paper is organised in the following manner. In Sec. \ref{sec:LPSC} we define the methodology concerning the propagation time delay of photon in Schwarzschild spacetime which we compared with the time delay in post-Newtonian while in Sec. \ref{sec:naked singularity} we apply our methodology in JNW and JMN-1 spacetimes and find the core results of our analysis. In Sec. \ref{sec:results} we discuss the results and their observational implications.

\section{\label{sec:LPSC}Light Propagation time in Schwarzschild Spacetime}
\begin{figure*}[htbp]
    \centering
    \includegraphics[width=0.98\linewidth]{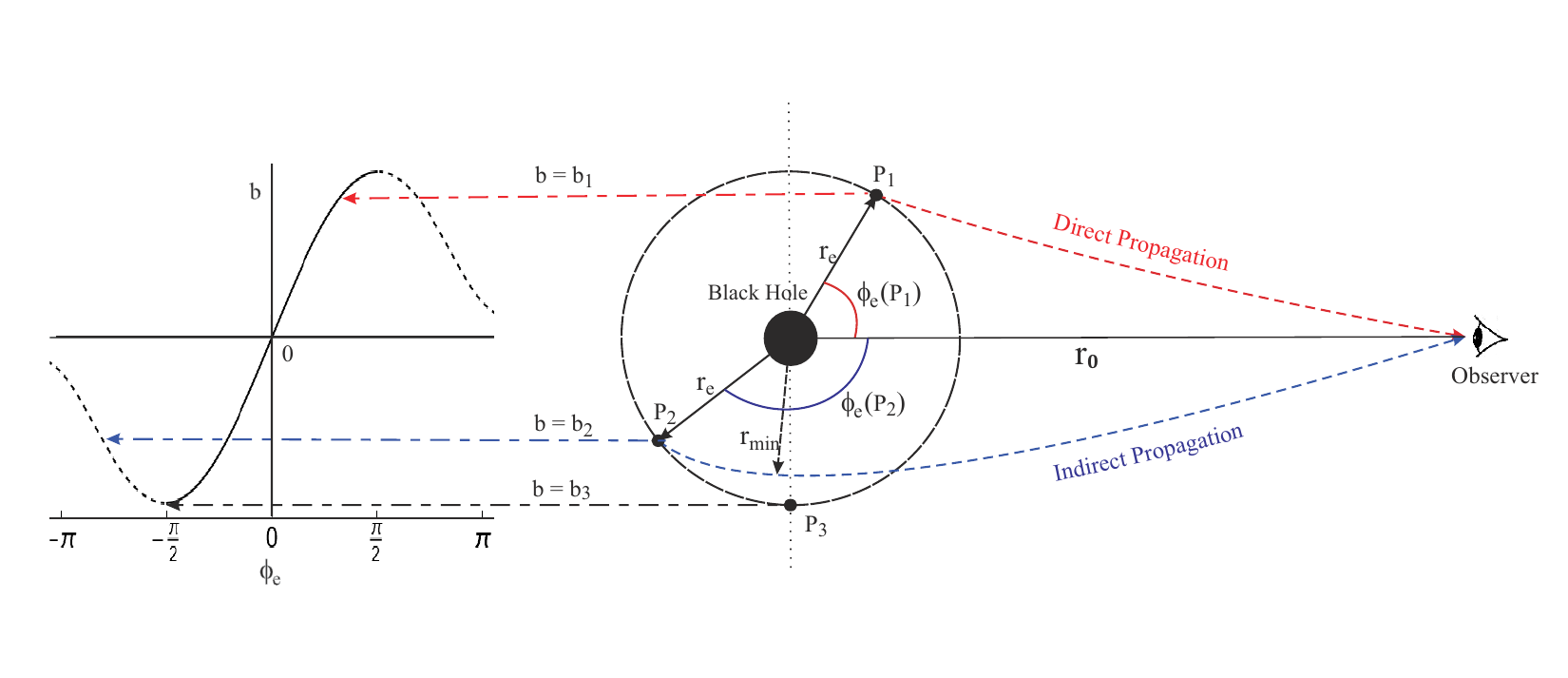}
    \caption{A schematic diagram illustrating the emitter-observer problem for a circular orbit of the pulsar on the equatorial plane. Pulsar positions given by the polar coordinates $(r_e,\phi_e)$ are denoted by $P_1$, $P_2$, and $P_3$ for different positions in the orbit of a black hole at radius $r_e$ shown by a long-dashed line. The observer is located at $(r_0,\phi_0)$, where $\phi_0$ equals zero. The diagram shows direct (red dashed line) and indirect (purple dashed line) photon propagation scenarios, depicting how photons reach the observer from the pulsar. On the left side of the figure, photon impact parameter variations with pulsar positions are represented using a three-dashed arrow line on ($b \rightarrow \phi_e$) plot.}
    \label{fig:schdia}
\end{figure*}

The Schwarzschild metric in Boyer–Lindquist coordinates can be written as 
\begin{align}
    \begin{split}
        ds^2=-\Big(1-\frac{2 M}{r}\Big)(cdt)^2
        &+\Big(1-\frac{2 M}{r}\Big)^{-1} dr^2\\
        &+r^2(d\theta^2+sin^2\theta d \phi^2) \;,
    \end{split}
\end{align}
where $M=\frac{Gm}{c^2}$ and $m$ is the mass of the black hole. While $G$ and $c$ are gravitational constant and speed of light respectively. The null geodesics along which photons propagate obey the geodesic equation
\begin{equation}
    \ddot{x}^\mu +\Gamma^\mu_{\nu\rho}\dot{x}^{\nu}\dot{x}^\rho=0 \;, 
\end{equation}
where $\Gamma$ denotes the Christoffel symbol, and the dot signifies the derivative with respect to the affine parameter $\tau$ along the curve. To retain the spherical symmetry of the spacetime, we set $\theta=\pi/2$ by considering the orbital plane as the equatorial plane. Consequently, we define the energy $E$ and the angular momentum $L$ as 
\begin{equation}
     E=g_{tt} (c\dot{x}^t) = -\Big(1-\frac{2M}{r}\Big) \frac{cdt}{d\tau} \;,
     \label{energy}
\end{equation}
\begin{equation}
    L=g_{\phi\phi}\dot{\phi}=r^2 \frac{d\phi}{d\tau} \;,
    \label{angular momentum}
\end{equation}
also the condition for null geodesics,
\begin{equation}
    g_{\mu\nu}\dot{x}^\mu \dot{x}^\nu=0 \;.
    \label{nullgeodesics}
\end{equation}
From Eqs. (\ref{energy}), (\ref{angular momentum}), and (\ref{nullgeodesics}) we can find the equations of motion for light,
\begin{align}
\begin{split}
    \bigg(\frac{dr}{d\tau} \bigg)^2 &=\frac{1}{g_{rr}} \bigg( \frac{E^2}{g_{tt}} -\frac{L^2}{g_{rr}}\bigg) \\
    &=\frac{L^2}{g_{rr}}\bigg(\frac{1}{b^2 g_{tt}}-\frac{1}{g_{\phi\phi}}\bigg)=L^2  R(r,b) \;,
\end{split}
\\
    \frac{d\phi}{d\tau}&=\frac{L}{g_{\phi\phi}}=\frac{L}{r^2} \;,
\\
    \frac{cdt}{d\tau}&=\frac{E}{g_{tt}}=\frac{E}{1-\frac{2M}{r}} \;,
\end{align}
where, the impact parameter $b=L/E$ and
\begin{align}
    \begin{split}
             R(r,b)=\frac{1}{r^4}\bigg(\frac{r^4}{b^2}-r^2+2Mr\bigg) \;.
     \label{polynomial}
    \end{split}    
\end{align}
To locate the turning point of the null geodesic, we set the component $dr/d\tau =0$ and determine the roots of $R(r,b)$. The quartic polynomial $R(r,b)$ possesses four roots: one at zero and one as a negative root for all values of $b$. The other two roots vary based on the impact parameter $b$. When $b=b_{crit}=\sqrt{27} \text{M}$, these roots combine to yield a single positive root, indicating an unstable circular orbit. For $b>b_{crit}$, two positive roots emerge, resulting in either a flyby orbit (from infinity to the closest approach and then to the observer) or a terminating bound orbit. However, for $b<b_{crit}$, a complex pair arises, resulting in a terminating escaping orbit (from infinity and falls into the singularity at $r=0$). \cite{Hackmann:2018fmk}.

Consider a pulsar orbiting the GC in a circular orbit as shown in Figure \ref{fig:schdia}. Given the substantial mass distinction between the GC and the pulsar, the pulsar is treated as a test particle emitting radio pulses at different positions in the orbit shown as a circular long-dashed line. We are interested in finding the arrival time of the photons detected by an observer situated at a large distance $r_0$. Therefore, we focus on the flyby orbit, where a photon emitted from the pulsar either reaches the observer directly shown by a red dashed line in Figure \ref{fig:schdia} or travels to the closest approach ($r_{min}$), the largest root of the polynomial in equation (\ref{polynomial}) before directed to the observer shown as a purple dashed line. The trajectory $\gamma$ is contingent upon the pulsar's position. If the pulsar is in front of the GC from the observer's frame ($P_1$), it indicates direct propagation. Conversely, if the pulsar is behind the GC in the observer's frame ($P_2$), it signifies indirect propagation.

To determine the arrival time of the photon, we are required to solve the emitter-observer problem in order to find the impact parameter $b$ of corresponding propagating photons determined as a grid of pulsar positions in the circular orbit. We also need to match direct and indirect propagation at the pulsar position $P_3$  in order to find the impact parameter. From equations of motion, we get
\begin{align}
    \frac{dr}{d\phi} &=g_{\phi\phi}\sqrt{R(r,b)}\; ,\\
    \Delta \phi &= \int_{\gamma}  \frac{dr}{g_{\phi\phi}\sqrt{R(r,b)}} {\label{emmiter-observer}}\;,
\end{align} 
where $\Delta \phi$ is the angle difference along the photon trajectory $\gamma $ in the instantaneous common plane of pulsar and observer. There is no exact analytical solution to equation (\ref{emmiter-observer}) is available. However, there is some literature in which authors have tried to approximate the analytical solution for the Schwarzschild black hole \cite{Semerak:2014kra, Beloborodov:2002mr, DeFalco:2016yox}. The approximation methods are not sufficient enough for our study, and hence we solve it completely numerically for both direct and indirect propagation. In direct propagation, the photon travels directly from $r_e$ to $r_o$. On the other hand, in indirect propagation, the photon has to travel an additional distance. The photon first goes to the minimum approach and then to the observer ($r_e \rightarrow r_{min} \rightarrow r_o$). Hence, we are left with 
\begin{align}
    \Delta \phi _{d} (SC) &= \int_{r_e}^{r_{o}}  \frac{dr}{r^2\sqrt{R(r,b)}} \; ,\\
    \Delta \phi _{id}(SC) &= \bigg[\int_{r_e}^{r_{min}} +\int_{r_{min}}^{r_o} \bigg] \frac{dr}{r^2\sqrt{R(r,b)}} \; .
\end{align}
For this analysis, we consider Sgr-A* with the mass $M = 4 \times 10^{6} M_{\odot}$ and $M_{\odot} = \frac {G m_{\odot}}{c^2} = 1476 \text{m}$, where $m_{\odot}$ is mass of the sun, along with an Earth-based observer situated at a distance of approximately $8$ kpc, equivalent to $4 \times 10^{10} \text{M}$  from the Sgr-A*. Assuming the pulsar orbits in a circular path at a distance of $100 \text{M}$ from the center. These conventions are taken from \cite{Hackmann:2018fmk}.\\
Also, we need to find the largest value of impact parameter $b_{max}$ at the separation point $\pi/2$ and $-\pi/2$. We achieve this by finding the root of $b$ in $dr/d\tau = 0$ at $r=r_e$. The emitter-observer problem is then solved numerically for equation (\ref{emmiter-observer}).
\begin{figure}
    \centering
    \vspace{0.2cm}
    \includegraphics[width=1\linewidth]{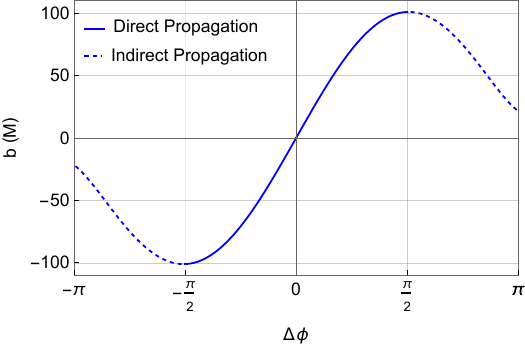}
    \caption{Emitter-observer problem for Schwarzschild metric for a fixed observer located at $r_0 = 8 \text{kpc}$. The plot represents values of the impact parameter ($b$) for $\Delta\phi \in [-\pi, \pi]$. The dashed line represents indirect propagation while the solid lines show direct propagation. }
    \label{fig:emitter-observer}
\end{figure}\\

Figure~\ref{fig:emitter-observer} represents the solution of the emitter-observer problem. The separation point is slightly diverted from $\pi/2$ and $-\pi/2$ due to the strong gravitational lensing effect. For a sufficiently larger orbit, the separation is smaller, and vice versa. When the pulsar is positioned directly behind the ultra-compact object at $\Delta\phi=\pi$ and $-\pi$,  there is no actual discontinuity in the observations. It is plotted in the positive and negative direction of the impact parameter for easier visualization as given in \cite{DellaMonica:2023ydm}. Since the impact parameter is always positive, it smoothly matches at $\Delta\phi = \pi$ and $\Delta\phi = -\pi$.

Once we have the impact parameter ($b$) associated with the angle difference of photon between emission and observation ($\Delta \phi$), our focus shifts to determining the arrival time of photons. For simplicity, we initially disregard the gravitational field of the Sun, as well as the Earth's motion around it. Nonetheless, it's possible to incorporate these disregarded effects later by employing a weak field approximation for the Sun \cite{Damour:1986}.

From equations of motion,
\begin{align}
    \frac{dt}{dr} &= \frac{1/b}{g_{tt} \sqrt{R(r,b)}} \; , \quad \\
    \Delta t  \equiv  t_o -t_e &= \int_{r_e,\gamma}^{r_o} \frac{1/b}{g_{tt}\sqrt{R(r,b)}} dr \;.
    \label{arrivaltime}
\end{align}

We again define two paths for photons to follow, direct and indirect corresponding to the solution of the position of the pulsar in the range of $[-\pi,\pi]$. During $-\pi/2$ to $\pi/2$, the direct propagation is considered under the access of impact parameter solved from the emitter-observer problem. Similarly for $-\pi$ to $-\pi/2$ and $\pi/2$ to $\pi$, indirect propagation,
\begin{equation}
     \Delta t_{d} (SC) = \int_{r_e}^{r_o} \frac{1}{b (1- \frac{2M}{r}) \sqrt{R(r,b)}}\ dr \;,
\end{equation} 
\begin{equation}
    \begin{split}
        \Delta t_{id}(SC) =\int_{r_e}^{r_{min}} \frac{1}{b (1- \frac{2M}{r}) \sqrt{R(r,b)}} dr \\
     + \int_{r_{min}}^{r_o} \frac{1}{b (1- \frac{2M}{r}) \sqrt{R(r,b)}} dr \;.
    \end{split}
\end{equation}
The arrival time has no analytical solution available. However, in \cite{Hackmann:2018fmk} the authors try to solve the integral for the Schwarzschild spacetime using the Jacobian elliptical integral to find the exact solution. Here, we solve the integral numerically and hence do not require the use of Jacobian elliptical functions. 

Considering a Keplerian orbit of the pulsar (The assumption has no impact on our method for the fully relativistic solution), we have $r_e = \frac{a (1- e^2)}{1+e\, cos\,\phi_e}$ where $a$ is semi-major axis, e is eccentricity and $\phi_e$ is true anomaly. From \cite{Hackmann:2018fmk} we adopt a model for our calculation with an inclination angle $i=\pi/3$ in a common plane along with the argument of periastron $\omega = -\pi/2$. Therefore, the angle $\phi$ in equation (\ref{emmiter-observer}), in the instantaneous common plane of pulsar and observer, is determined as 
\begin{equation}
    cos \,\phi = -sin\,i\, \times sin(\omega+\phi_e) \; . 
    \label{true anomaly}
\end{equation}
Using equation (\ref{true anomaly}) we can now find the true anomaly $\phi_e$ from the propagation angle $\phi$ in a common plane which is dependent on the impact parameter from the emitter-observer. In our model since we are using pulsar orbit with zero eccentricity, the true anomaly becomes the mean anomaly.

Now that we have the mean anomaly of the pulsar's circular orbit as a function of the impact parameter, we can proceed to find the arrival time. Here we use a methodology in which we find the numerical value of the minimum approach at the grid of points between $[-\pi,\pi]$ to describe the motion for direct and indirect propagation then interpolating all the discrete values in the arrival time formula. The whole arrival time problem is solved solely with the value of the impact parameter hence we define every function with only the impact parameter as a variable function including $r_{min}$. Also, to verify our methodology we subtract the R\"oemer time and Shapiro time as post-Newtonian approximation from the derived Schwarzschild spacetime expression according to \cite{Hackmann:2018fmk}. Here, note that we define the reference point at $\pi/2$.
\begin{figure}[htbp]
    \centering
    \includegraphics[width=1\linewidth]{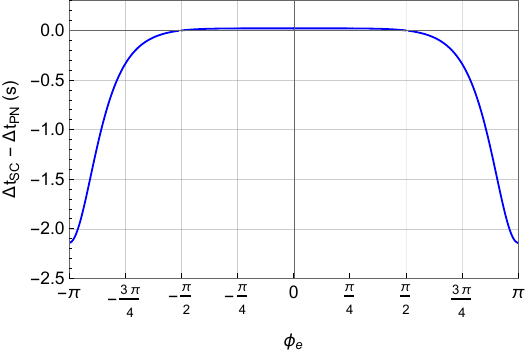}
    \caption{Time delay difference in Schwazschild spacetime and post-Newtonian approach in seconds from $-\pi$ to $\pi$ pulsar positions for the model considered using direct and indirect propagation formulas derived.}
    \label{fig:PN}
\end{figure}

In Figure~\ref{fig:PN}, we calculated the time delay discrepancy between Schwarzschild spacetime and post-Newtonian, which is already explored in \cite{Hackmann:2018fmk} to validate our approach. We can see that during direct propagation, at $\phi_e=0$ the time delay is $21.4 ~\text{ms}$, however, the main difference can be seen in indirect propagation, where the difference is $-2.137~\text{s}$ at $\phi_e = \pi$ and $\phi_e = -\pi$ which is the exact back side of the black hole. This verifies our results by matching the analysis carried out in \cite{Hackmann:2018fmk}. 

Since we are currently using only one orbit at $r_e=100 M$, the other orbits at different radii would show different behaviours. When the pulsar orbit is closer to the GC, the time delay difference between Schwarzschild spacetime and post-Newtonian would increase due to the extreme geometrical effect near the GC. However, when the orbit is farther away from the GC it shows less difference.

\section{\label{sec:naked singularity}Propagation time delay in naked singularities}

After formulating the methodology of finding the time delay in Schwarzschild spacetime we can use the same formulation for alternative spacetimes as well. In \cite{DellaMonica:2023ydm} authors have derived the time delay difference between the Schwarzschild and its alternative black hole spacetimes. As the behaviour of the spacetime metric affects the null geodesics we now apply the formulation to find the arrival time in naked singularity spacetimes and compare it with the Schwarzschild spacetime. The naked singularity is a horizonless singularity from which the past incomplete null geodesics can exit from the singularity. There are two naked singularity models we have considered in this paper as follows.\\

\subsection{Janis-Newman-Winicour Spacetime}

The first spacetime we studied was given by Janis, Newman, and Winicour in 1968 \cite{jnw} with a massless scalar field. The JNW solution satisfies the weak energy condition and has a strong globally naked singularity \cite{Virbhadra:1995iy}. The metric of JNW can be written as  
\begin{multline}
    ds^2 = - \bigg(1-\frac{2\mathcal{M}}{r}\bigg)^\gamma dt^2 + \bigg(1-\frac{2\mathcal{M}}{r}\bigg)^{-\gamma} dr^2 
    \\+ \bigg(1-\frac{2\mathcal{M}}{r}\bigg)^{1-\gamma}r^2 d\Omega^2 \;,
    \label{JNW eq}
\end{multline}
with $d\Omega^2 = d\theta^2 + sin^2\theta\, d\phi^2$ and associated scalar field, 
\begin{align}
    \Phi(r)=\frac{\sqrt{1-\gamma^2}}{2} \, ln\bigg(1-\frac{2\mathcal{M}}{r}\bigg) 
    \\= \frac{q}{2\mathcal{M}} \, ln\bigg(1-\frac{2\mathcal{M}}{r}\bigg) \;,
\end{align}
where variable $\mathcal{M}$ is related to the ADM mass $\text{M}$ of the central singularity and the scalar charge $\text{q}$. $\mathcal{M}$ can be defined as 
\begin{equation}
    \mathcal{M}=\sqrt{M^2+q^2}\; ,\quad \gamma=\frac{M}{\mathcal{M}} \;,
\end{equation}
where the scalar charge can be formulated as
\begin{equation}
    q = M \bigg(\frac{1-\gamma^2 }{\gamma^2}\bigg) \;,
\end{equation}
with $0<\gamma<1 $. We can obtain the Schwarzschild spacetime solution just by putting $q=0$ in equation (\ref{JNW eq}). For simplicity, we take the equatorial plan, hence, $\theta=\pi/2$ and $\phi=0$. Using the condition of null geodesics from equation (\ref{nullgeodesics}), we can find the equations of motion for photon in the JNW metric.
\begin{align}
\begin{split}
    \Big(\frac{dr}{d\tau} \Big)^2&=L^2\bigg[\frac{1}{b^2}-\frac{1}{\big(1-\frac{2\mathcal{M}}{r}\big)^{1-2\gamma}r^2}\bigg]\\
    &=L^2 R_{JNW}(r,b) \;,
\end{split}
\\
    \frac{d\phi}{d\tau}&=\frac{L} {g_{\phi\phi}}=\frac{L}{\big(1-\frac{2\mathcal{M}}{r}\big)^{1-\gamma}r^2} \;,
\\
    \frac{cdt}{d\tau}&=\frac{E}{g_{tt}}=\frac{E}{\big(1-\frac{2\mathcal{M}}{r}\big)^\gamma} \;.
\end{align}

We solve the largest root of $R_{JNW}(r,b)$ for different values of impact parameter $b$ and specific value of scalar change $q$. For the JNW case, we use the same approach shown in Figure \ref{fig:schdia}. Here, instead of a black hole at the center, we have a naked singularity and solve the emitter-observer problem from equation (\ref{emmiter-observer}).
\begin{equation}
    \Delta\phi_{JNW}=\int_{\gamma} \frac{dr}{\big(1-\frac{2\mathcal{M}}{r}\big)^{1-\gamma}r^2 \sqrt{R_{JNW}(r,b)}} \;.
    \label{emitterobserverjnw}
\end{equation}

In this equation, we take different paths according to the direct and indirect propagation and solve the emitter-observer problem.\\
After solving the emitter-observer problem as a function of impact parameter $b$ we now move forward to find the arrival time for direct as well indirect propagation for the JNW using equation (\ref{arrivaltime}). 
\begin{equation}
    \Delta t_{d} (JNW)= \int_{r_e}^{r_0} \frac{dr}{b \big(1-\frac{2\mathcal{M}}{r}\big)^\gamma \sqrt{R_{JNW}(r,b)} } ,
\end{equation}
\begin{equation}
\begin{split}
    \Delta t_{id} (JNW) = \int_{r_e}^{r_{min}} \frac{dr}{b \big(1-\frac{2\mathcal{M}}{r}\big)^\gamma \sqrt{R_{JNW}(r,b)} } 
    \\
    +\int_{r_{min}}^{r_0} \frac{dr}{b \big(1-\frac{2\mathcal{M}}{r}\big)^\gamma \sqrt{R_{JNW}(r,b)} }  .
\end{split}
\end{equation}
\begin{figure}[htbp]
    \centering
    \includegraphics[width=1\linewidth]{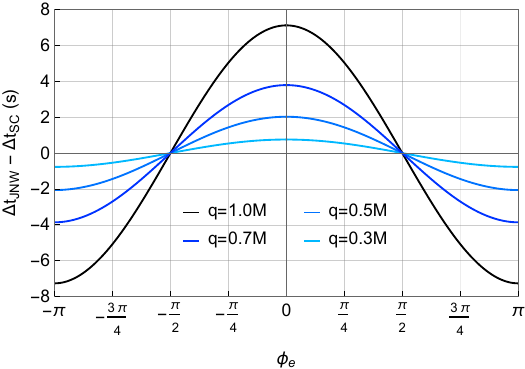}
     \caption{The time delay in JNW naked singularity in comparison with the Schwarzschild black hole for the different values of scalar charge ($q$). The considered mass of central object Sgr-A* is $M=4 \times 10^6 M_{\odot}$. While the pulsar is in the $r_e =100\text{M}$ circular inclined orbit at $i=\pi/3$ with the earth-based observer at $r_0 =8kpc$ from the GC.}
    \label{fig:JNW}
\end{figure}

\begin{figure*}[htbp]
    \centering
    \includegraphics[width=1\linewidth]{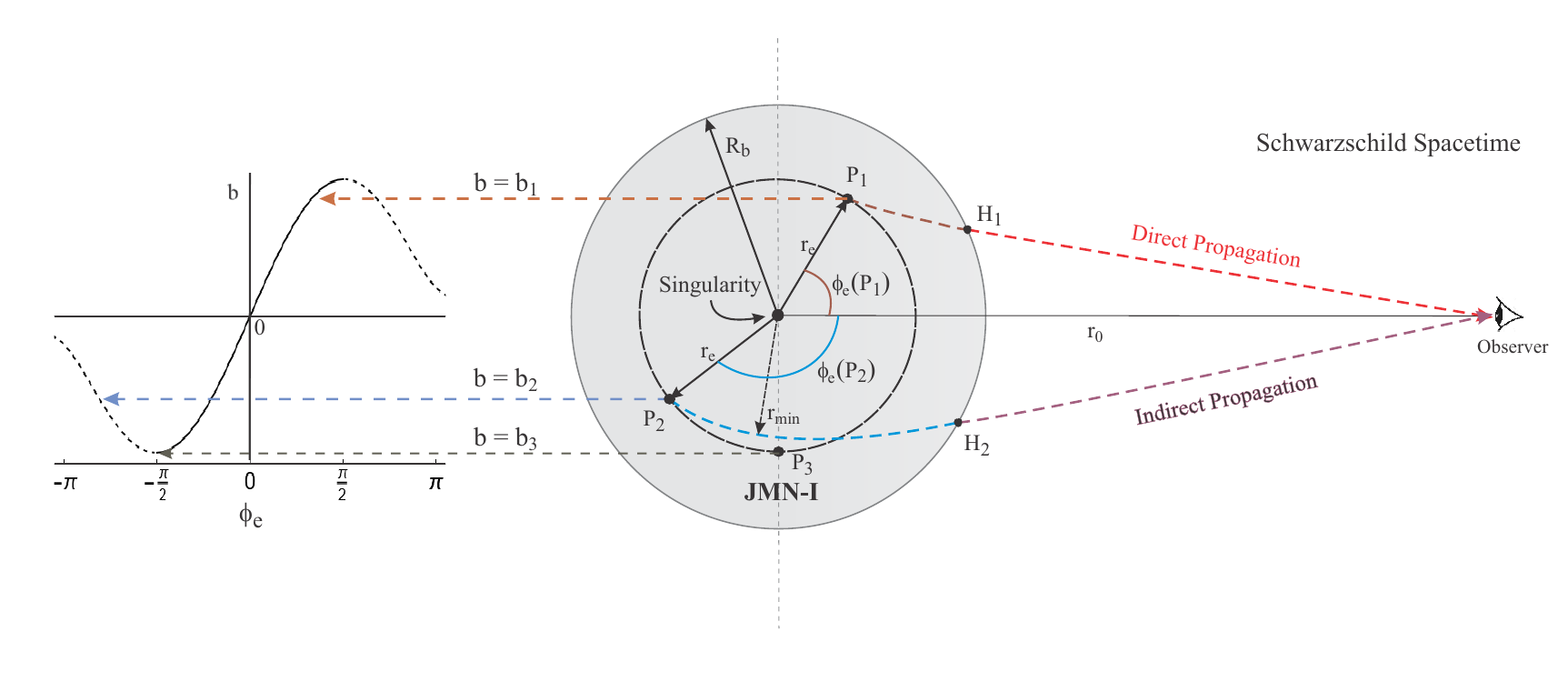}
    \caption{The figure illustrates the emitter-observer problem in JMN-1 spacetime. The JMN-1 spacetime is enclosed inside of the $R_b$ radius (shaded portion) with a singularity at the center and outside is the Schwarzschild spacetime with matching at $R_b$. The pulsar orbit (long-dashed line) is inside of the JNW-1 spacetime. In the photon paths, we also matched both spacetimes at points denoted by $H_1$ and $H_2$ for direct (brown and red dashed line) and indirect (blue and purple dashed line) propagation respectively. The left side of the figure shows the impact parameter for the photon corresponding the pulsar position $(r_e, \phi_e)$ shown by a three-dashed arrow line indicating on ($b\rightarrow \phi_e $) plot. }
    \label{fig:JMNsche} 
\end{figure*}

In Figure~\ref{fig:JNW} we examine the time delay of radio pulses in JNW spacetime for a pulsar in a circular orbit in comparison with the Schwarzschild black hole. The scalar charge takes values between $0 ~\text{M}$ and $ 1~\text{M}$. As shown in Figure~\ref{fig:JNW}, for $q=1~\text{M}$ (indicated by the solid black line), the time delay when compared to Schwarzschild spacetime reaches up to $-7.27 ~\text{s}$ at $\phi_e = \pi$ and  $-\pi$, whereas at $\phi_e = 0$ the difference is $7.13 ~\text{s}$ which is total of $14.40 \text{s}$. We analysed that a smaller scalar charge yields a closer match to the Schwarzschild spacetime model. 

When we compare the Schwarzschild spacetime with the post-Newtonian approximation, we have focused on identifying the quality of the post-Newtonian approximation (only accounting for Roemer and Shapiro delay) \cite{Hackmann:2018fmk}, which is why the difference in direct and indirect photon propagation time delay difference was observed. Since we are only comparing the time delay of the signal between the JNW and Schwarzschild spacetimes, it looks symmetric and the deference increases for larger values of the scalar field conserving the symmetric behaviour.

\subsection{ Joshi-Malafarina-Narayan Spacetime}

The second spacetime we use was given by Joshi, Malafarina, and Narayan \cite{psj2} which is a result of an inhomogeneous dust collapse and a better alternative to the OSD collapse \cite{Oppenheimer:1939ue, Datt:1938} as the OSD collapse model assumes an idealistic set of initial conditions for a homogeneous dust collapse. The solution holds a time-like naked singularity at $r=0$ and does not allow the formation of trapped surfaces. 
The metric is given by

\begin{equation}
    ds^2=-(1-M_0 )\bigg(\frac{r}{R_b} \bigg)^\frac{M_0}{1-M_0} dt^2+\bigg(\frac{1}{1-M_0}\bigg)dr^2 +r^2 d\Omega^2 ,
\end{equation}
where $M_0$ and $R_b$ are positive constants and $R_b$ represents the boundary radius of the distributed matter around a central singularity of JMN-1 spacetime. The parameter $M_0$ is defined in the range $0 < M_0 < 4/5$, where the upper limit of $M_0$ is defined by the fact that the sound speed cannot exceed unity \cite{psj2}. For modeling of the isolated objects, spacetime should be asymptotically flat. In this present work, we have considered JMN-1 spacetime to model astrophysical compact object where Schwarzschild spacetime is the best exterior solution. For cosmological scenarios, one can induce some sort of cosmological constant to get the asymptotic flatness. The total mass $M$ is given by $M=\frac{1}{2}M_0 R_b$. The metric is naturally matching with the Schwarzschild spacetime at an outer radius at $r=R_b$. 

From null geodesics, we get the following equations of motion 
\begin{align}
\begin{split}
    \Big(\frac{dr}{d\tau} \Big)^2&= L^2 \bigg[\frac{1}{b^2}\bigg(\frac{R_b}{r}\bigg)^{\frac{M_0}{1-M_0}} -\frac{1-M_0}{r^2} \bigg] \\
    &=L^2 R_{JMN}(r,b) ,
\end{split}
\\
    \frac{d\phi}{d\tau}&=\frac{L}{r^2} ,
\\
    \frac{cdt}{d\tau}&=\frac{E}{1-M_0}\bigg(\frac{R_b}{r}\bigg)^{\frac{M_0}{1-M_0}} .
\end{align}

We again solve the emitter-observer problem which is shown in Figure \ref{fig:JMNsche} using equation (\ref{emmiter-observer}) for JMN-1 and find the angle difference between the emission of the photon from a pulsar path and the observer at the Earth which we later connect with the impact parameter using the following equation derived from equations of motion 

\begin{equation}
    \Delta\phi(JMN)=\int_{\gamma}\frac{dr}{r^2 \sqrt{R_{JMN}(r,b)}}. 
\end{equation}

Since JMN-1 is not asymptotically flat, we considered the JMN-1 as interior spacetime inside the radius $R_b$ and the pulsar orbit within the sphere $0<r_e<R_b$. Outside of the sphere, the exterior Schwarzschild spacetime is matched for asymptotic flatness at large distances where an Earth-based observer resides. Hence, it is required to match the null geodesics path for JMN-1 and Schwarzschild spacetime at $r=R_b$.
To properly match the spacetimes we need to match the impact parameters of both spacetimes. Hence we match $R(r,b)$ for different values of $r$ and find $b$ for Schwarzschild spacetime which we now denote as $b_{sc}$. 
The time delay can be found from 
\begin{equation}
\begin{split}
    \Delta t_{d} (JMN) &= \int_{r_e}^{R_b} \frac{(R_b/r)^{(M_0/{1-M_0})} \, dr}{b \big(1-M_0\big) \sqrt{R_{JMN}(r,b)}}
    \\
    &+ \int_{R_b}^{r_o} \frac{dr}{b_{sc} (1- \frac{2M}{r}) \sqrt{R_{sc}(r,b_{sc})}} \;,
\end{split}
\end{equation}
\begin{equation}
\begin{split}
    \Delta t_{d} (JMN) &= \int_{r_e}^{r_{min}} \frac{(R_b/r)^{(M_0/{1-M_0})} \, dr}{b \big(1-M_0\big) \sqrt{R_{JMN}(r,b)}}
    \\
    &+ \int_{r_{min}}^{R_b} \frac{(R_b/r)^{(M_0/{1-M_0})} \, dr}{b \big(1-M_0\big) \sqrt{R_{JMN}(r,b)}}
    \\
    &+ \int_{R_b}^{r_o} \frac{dr}{b_{sc} (1- \frac{2M}{r}) \sqrt{R_{sc}(r,b_{sc})}}  \; .
\end{split}
\end{equation}

\begin{figure}[htbp]
    \centering
    \includegraphics[width=0.98\linewidth]{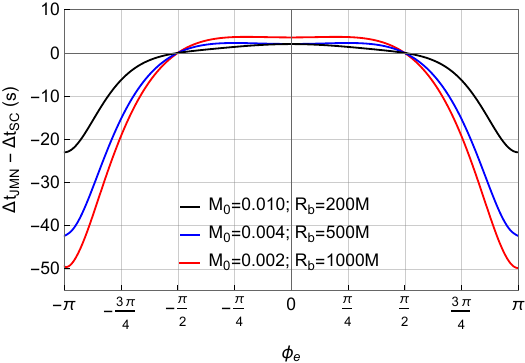}
    \caption{The time delay in JMN-1 compared with the Schwarzschild spacetime for the different model parameters $M_0$ and $R_b$. The mass of the GC, circular pulsar orbit radius, and inclination is $M=4 \times 10^6 M_{\odot}$, $r_e = 100 \text{M}$, and $i=\pi/3$ respectively. Where model parameter $R_b$ value is always larger than the pulsar orbit. In the region of $R_b<r<r_0$, the Schwarzschild metric is considered to achieve the asymptotic flatness for the earth-based observer.} 
    \label{fig:JMN-1}
\end{figure}

As shown in Figure~\ref{fig:JMN-1} we assume different values of model parameters to find out the time delay in JMN-1 in comparison to Schwarzschild spacetime. We used three different values of the model parameter $M_0$ and found the difference. It can be observed that when $M_0 = 0.002$, which corresponds to $R_b = 1000 \text{M}$, we find a very large time delay difference. At $\phi_e = \pi$ and  $-\pi$, the propagation time delay difference is almost $-50~\text{s}$. On the other hand, at $\phi_e = 0$, we get only $ \sim 3.6~\text{s}$. The 50-sec delay difference between JMN-1 and Schwarzschild spacetime is observed when considering a very large sphere of matter ($R_b=1000M$). This affects largely on photon path leading to a more significant difference in propagation times. However, when we consider the smaller matter field sphere of JMN-1 spacetime, the difference would be greatly reduced. We observe that as more photon traverse through JMN-1 spacetime, the less propagation time delay it experiences compared to the Schwarzschild black hole during indirect propagation. Conversely, in direct propagation, photons take more time in propagation compared to the Schwarzschild black hole.

Our study shows that these time delay differences can range from a few milliseconds to tens of seconds depending on the pedagogical parameter space, varying significantly across different spacetime geometries. These distinctive differences could enable future pulsar timing analysis around an ultra-compact object to not only constrain model parameters but also to rule out models that exhibit inconsistent propagation time delay patterns. For particularly in JMN-1 spacetime, the large time delay difference occurred for very small value of $M_0$ which is near to zero, while for $M_0>2/3$, this time delay difference is almost indistinguishable from the Schwarzschild black hole.

\section{\label{sec:results}Results and Discussion} 

Studying radio pulses generated by the pulsar in the proximity of the GC is the main motivation for this study. Observationally detecting a pulsar in a tight orbit of GC can open up unprecedented opportunities for studying extreme gravity physics, testing GR, and understanding the nature of the compact object. 

The rotating black holes or naked singularities likely better match the observational data for Sgr A* than non-rotating models. However, for simplicity and transparency, we've limited our study to static, spherically symmetric metrics, neglecting spin effects, which minimally impact the shadow radius \cite{Vagnozzi:2022moj}. The spin and inclination angle of Sgr A* remains uncertain, with EHT images consistent with both high and low spin scenarios, depending on the Kerr metric \cite{EventHorizonTelescope:2022xnr,EventHorizonTelescope:2022urf}. While various estimates suggest a broad range of spin values \cite{Broderick:2008sp,Kato:2009zw,Broderick:2011mk,Huang:2009sq,Shcherbakov:2010ki}, recent work indicates that Sgr A*’s spin might be very low \cite{Fragione:2020khu,Fragione:2022oau}. Given these inconsistencies, we adopt a conservative approach, neglecting spin effects while noting the possibility of low spin.

Hence, we numerically measured the propagation time delay of photons generated by the pulsar in a circular orbit in various static spacetimes assuming the source to be Sgr-A*, with mass $M=4 \times 10^6 M_{\odot}$ and Earth-based observer at a distance of $r_0 = 8kpc$. We first solved the geodesic equations of propagating photons to consider pure GR effects and numerically solved the emitter-observer problem shown in Figure \ref{fig:emitter-observer}. Once we get access to the impact parameter from the emitter-observer problem, we find the time delay in Schwarzschild spacetime. The post-Newtonian approximation is considered in Figure \ref{fig:PN} to identify the quality of the post-Newtonian approximation on the photons' trajectories. The time delay difference is measured to be $-2.137 ~\text{s}$ during the indirect proportion at $\phi_e = \pi $ and  $-\pi$. On the other hand, during the direct propagation, the delay difference is comparably very small around $21.4 ~\text{ms}$ at $\phi_e = 0$. 

The same formulation is then applied to JNW spacetime and compared with the Schwarzschild spacetime which is shown in Figure~\ref{fig:JNW}. We analyzed that the scalar charge is impacting largely on the behavior of photon propagation. For larger values of scalar charge ($q =1~\text{M}$) the measured time delay difference is very large. During direct propagation, the photon propagation in JNW takes approximately $7.13~\text{s}$ more compared to the Schwarzschild black hole at $\phi_e = 0$. On the other hand, during the indirect propagation, a photon traveling in JNW spacetime would take $7.27~\text{s}$ less at $\phi_e = \pi$ and  $-\pi$. Besides, in more realistic cases the massless scalar field would have a very small value of $`q`$ derived in \cite{Bambhaniya:2022xbz}, indicating an extremely small difference in time delay. 

On the contrary, JMN-1 spacetime is not asymptotically flat. Hence we considered a sphere of a certain radius ($R_b$) which has JMN-1 spacetime inside of it and outside is the Schwarzschild spacetime. To determine the time delay in JMN-1 spacetime, the orbit of the pulsar is considered inside the JMN-1 spacetime. We find out in Figure~\ref{fig:JMN-1} that the time delay difference is larger as compared to Schwarzschild spacetime and increases even more for a smaller value of $M_0$ (corresponding to a larger value of $R_b$). Besides, when the largest value of $M_0 = 0.01$ is considered ($R_b = 200\text{M}$), the time delay difference is very notable. At $\phi_e = 0$, it is measured to be around $2 ~\text{s}$ and at $\phi_e = \pi$ and  $-\pi$ it is $-23 ~\text{s}$ which extends up to $-50~\text{s}$ for very small value of $M_0 = 0.02$ ($R_b = 1000\text{M}$) at $\phi_e = \pi$ and  $-\pi$ while at $\phi_e = 0$ it is $3.6~\text{s}$. 

In this work, the spacetime geometries of naked singularities are not vacuum solutions of the Einstein field equations. Here, JMN-1 naked singularity represents an anisotropic fluid solution while JNW naked singularity represents a scalar filed solution, hence the causal structures of the spacetimes are different than the Schwarzschild spacetime. Therefore, due to non-vacuum surroundings in JMN-1 and JNW naked singularities, photons can have different propagation time. Now to factored our methodology within pulsar timing data analysis software like TEMPO \cite{Taylor:1989sw}, TEMPO2 \cite{Edwards:2006zg}, and PINT \cite{Luo:2020ksx}, we can compute high-precision timing residuals (differences between the observed time of arrival and those predicted by the model), and iteratively refine our models to achieve greater accuracy.

The future prospects of pulsar research are very promising and multifaceted. Pulsars' unique properties make them invaluable laboratories for studying stellar and binary evolution, as well as probing GR at a fundamental level. Recently, in a search of pulsars at our Milky Way GC, the EHT used 2017 observation data from its three most sensitive telescopes, the Atacama Large Millimeter/submillimeter Array, the Large Millimeter Telescope and the IRAM 30m Telescope but they did not find any significant pulsar signal due to insufficient sensitivity to detect pulsars \cite{EHT:2023hcj}. 

Furthermore, with upcoming high-precision observational facilities like the SKA \cite{Keane:2014vja}, which will reach sub-microsecond precision, and the use of Pulsar Timing Arrays (PTAs) for detecting low-frequency gravitational waves, incorporating such models can significantly enhance the study of pulsar dynamics, gravitational theories, and gravitational waves. These advancements make pulsar timing a crucial tool in astrophysics and fundamental physics research, providing a powerful way to probe the nature of spacetime in extreme environments. Also, the discovery of relativistic pulsar systems, potentially including pulsar-black hole binaries, holds the key to unraveling cosmic mysteries such as the cosmic censorship conjecture and the no-hair theorem. 

\section*{Acknowledgments}
The authors would like to thank Saurabh and the refree(s) for their valuable suggestions and comments.

\end{document}